\documentclass[a4,12pt]{iopart}
\usepackage{graphicx}

\begin{document}
\title{\bf{Empirical determination of charm quark energy loss and its
consequences for azimuthal anisotropy}}

\title[Empirical Determination......]{}
\author{Mohammed Younus \footnote{Email: mdyounus@vecc.gov.in} and
 Dinesh K. Srivastava \footnote{Email: dinesh@vecc.gov.in}}

\address{Variable Energy Cyclotron Center, 1/AF, Bidhan Nagar, Kolkata,
 700 064, India}

\begin{abstract}
We propose an empirical model to determine the
form of energy loss of charm quarks due to multiple scatterings
 in quark gluon plasma by demanding a good description of production of
D mesons and non-photonic electrons in relativistic collision of
heavy nuclei at RHIC and LHC energies. Best results
are obtained when we approximate the momentum loss per collision
$\Delta p_T = \alpha \, p_T$, where $\alpha$ is a constant depending on
the centrality and the centre of mass energy.
Comparing our results with those obtained earlier for drag coefficients
estimated using Langevin equation for heavy quarks we find that up to
half of the energy loss of charm quarks at top RHIC energy could be due to
collisions while that at LHC energy at 2760 GeV/A the collisional energy loss
could be about one third of the total.
Estimates are obtained for azimuthal anisotropy
in momentum spectra of heavy mesons, due to this energy loss.
We further suggest that energy loss of charm quarks may lead to an enhanced
production of D-mesons and single electrons at low $p_T$ in AA collisions.
\end{abstract}


PACS Indices: 14.65.Dw, 14.40.Lb, 13.30.Ce, 12.38.Mh, 25.75.Dw

\maketitle

\section{Introduction}

The vast amount of data collected at the Relativistic Heavy Ion
Collider at Brookhaven National Laboratory along with those
 recently collected from
the collision of lead nuclei at Large Hadron Collider at CERN
have led to the momentous discovery of quark gluon plasma (QGP).
A large part of the effort in the coming years will be devoted to
the determination of the precise values of the transport properties
of the QGP. In this context, the energy loss suffered by quarks of
 different flavours
and gluons as they traverse the QGP, undergoing collisions and radiating
gluons, is a subject matter of considerable topical interest.
 It is most simply demonstrated by a suppressed production of hadrons
 having large transverse momenta in nucleus-nucleus collisions when
compared to appropriately scaled productions in $pp$ collisions
at the same centre of mass energy per nucleon.

It is often suggested that heavy quarks may lose a smaller amount of energy
per unit length during their passage through QGP compared to light quarks,
due to the 'dead cone effect'~\cite{deadcone1,deadcone2}.
The experimental results, however,
show similar suppression for light and heavy mesons~\cite{suppression}.
 A recent calculation by
Abir et al~\cite{abir1} incorporating the generalized distribution of gluons
in $qc\rightarrow qcg$ and $gc\rightarrow gcg$ processes with
a proper accounting
of mass of the heavy quarks, leads to a $dE/dx$ for charm  quarks which is
quite similar to those for light quarks at energies $\geq $ 10-15 GeV.

The study of charm quark energy loss provides several unique advantages
over those of light partons. The charm mesons easily stand out in the
multitude of light mesons. Most of the charm quarks are produced in
initial fusion of quarks and anti-quarks
($q\overline{q}\rightarrow c\overline{c}$)
and gluons ($gg\rightarrow c\overline{c}$),
though a small additional production is expected~\cite{shor,lin,lmw,mdy,greiner}
from multiple scattering between jets, jets and thermalized partons,
and thermalized partons.
It is not yet clearly established if the charm quarks thermalize in
the QGP ~\cite{Bass},
though it is expected that due to their small numbers their impact on
 the bulk properties
of the QGP would be negligible. One also expects that due to their
large mass, charm
quarks will not change their direction as they traverse the plasma,
 though they will
slow down, making them excellent probes of azimuthal dependence
of the conditions of the interacting system.
As charm quarks have large mass, pQCD remains reasonably valid down to
lowest $p_T$.
There is one additional trait
which should help us in getting flavour dependence of energy loss of
heavy quarks. While a $u$ or a $d$ or a $s$ quark or a gluon can fragment
into one of many mesons or baryons, a charm or a bottom quark would
mostly fragment into only a $D$ or a $B$ meson or a charm or bottom baryon,
respectively.
Thus a charm quark after losing energy will appear as a D-meson having
lower energy. We shall see that this would lead to a characteristic
 enhancement of charm mesons or single electrons at low $p_T$.

The paper is organized as follows.
Sec.~2 contains expressions for calculations
of nuclear modification factor- $R_{AA}$, azimuthal anisotropy
coefficient- $v_2$,  charm production,
and our model prescription for energy loss. Sec.~3 contains
discussion of our results, followed by a summary in Sec.~4.

\begin{figure*}[h]
\begin{center}
\includegraphics[height=3in,width=3in,angle=270]{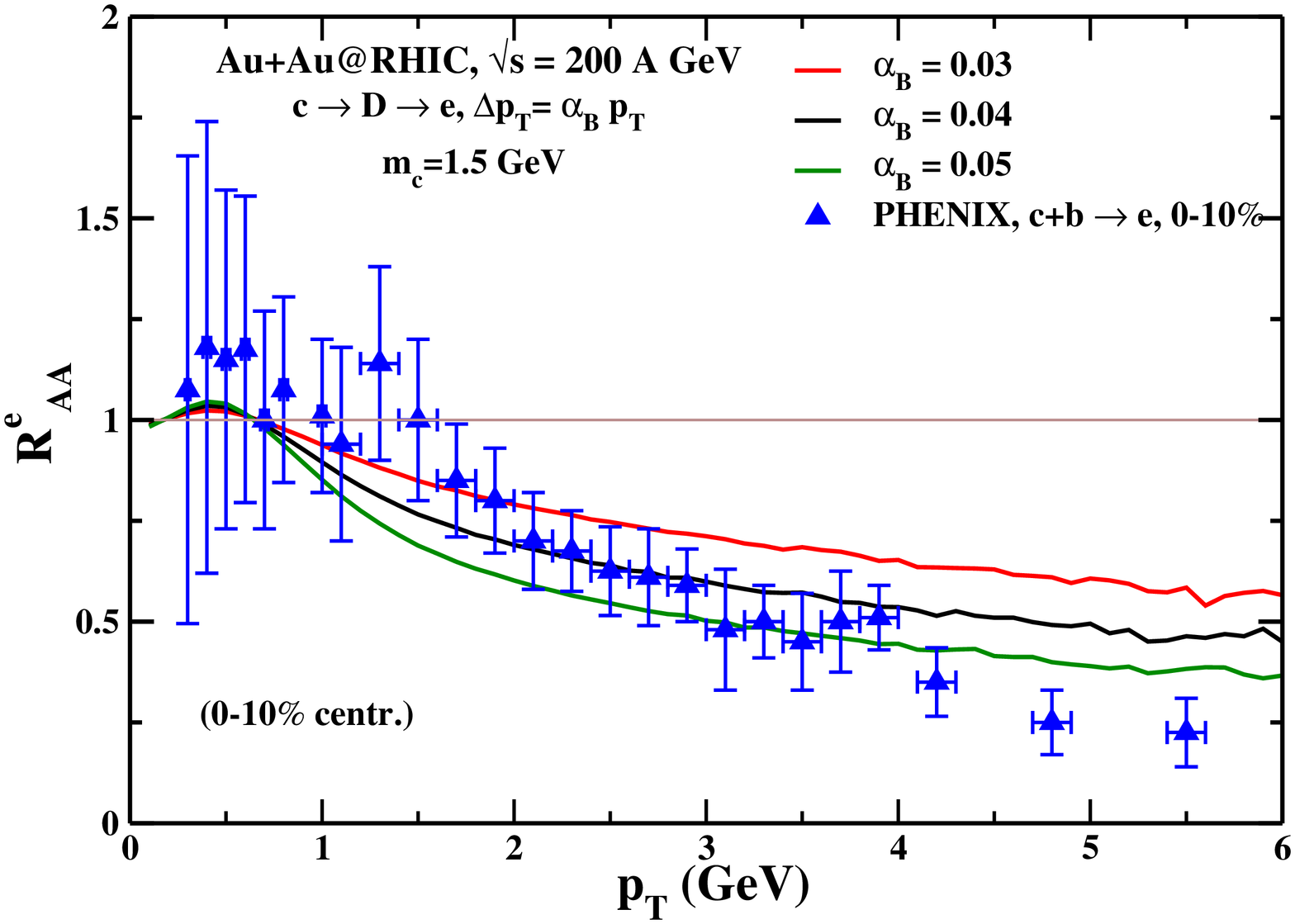}
\includegraphics[height=3in,width=3in,angle=270]{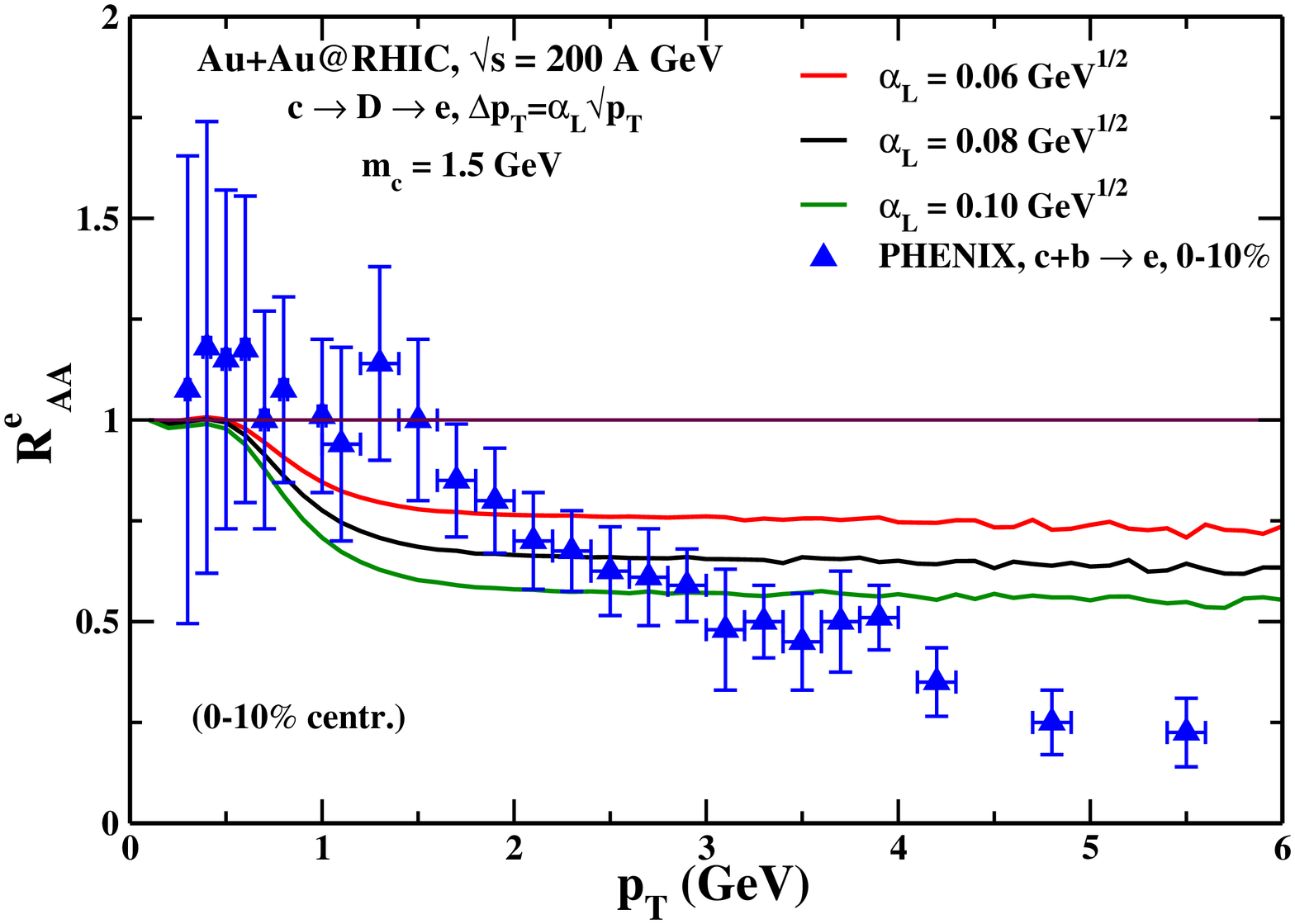}
\includegraphics[height=3in,width=3in,angle=270]{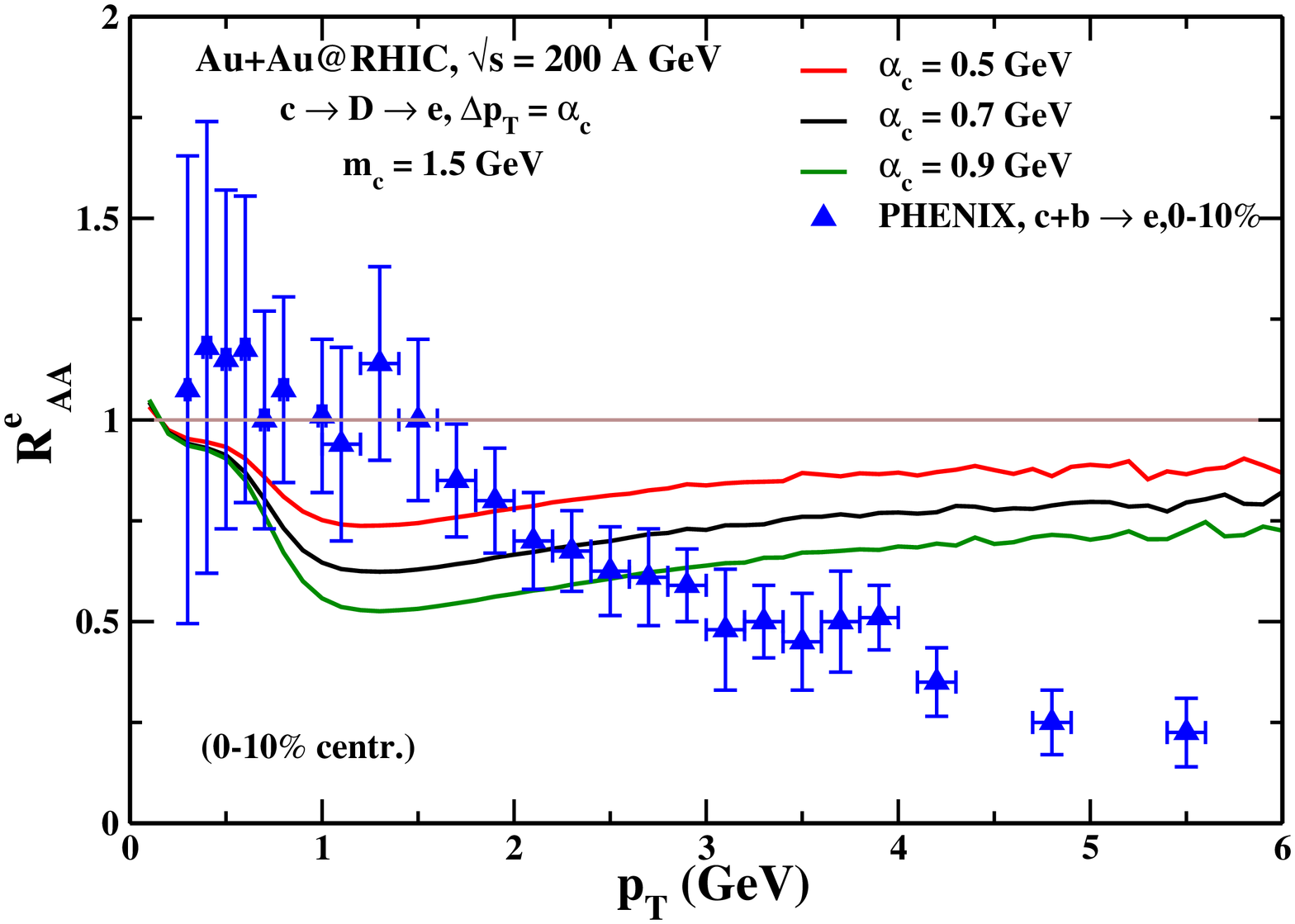}
\includegraphics[height=3in,width=3in,,angle=270]{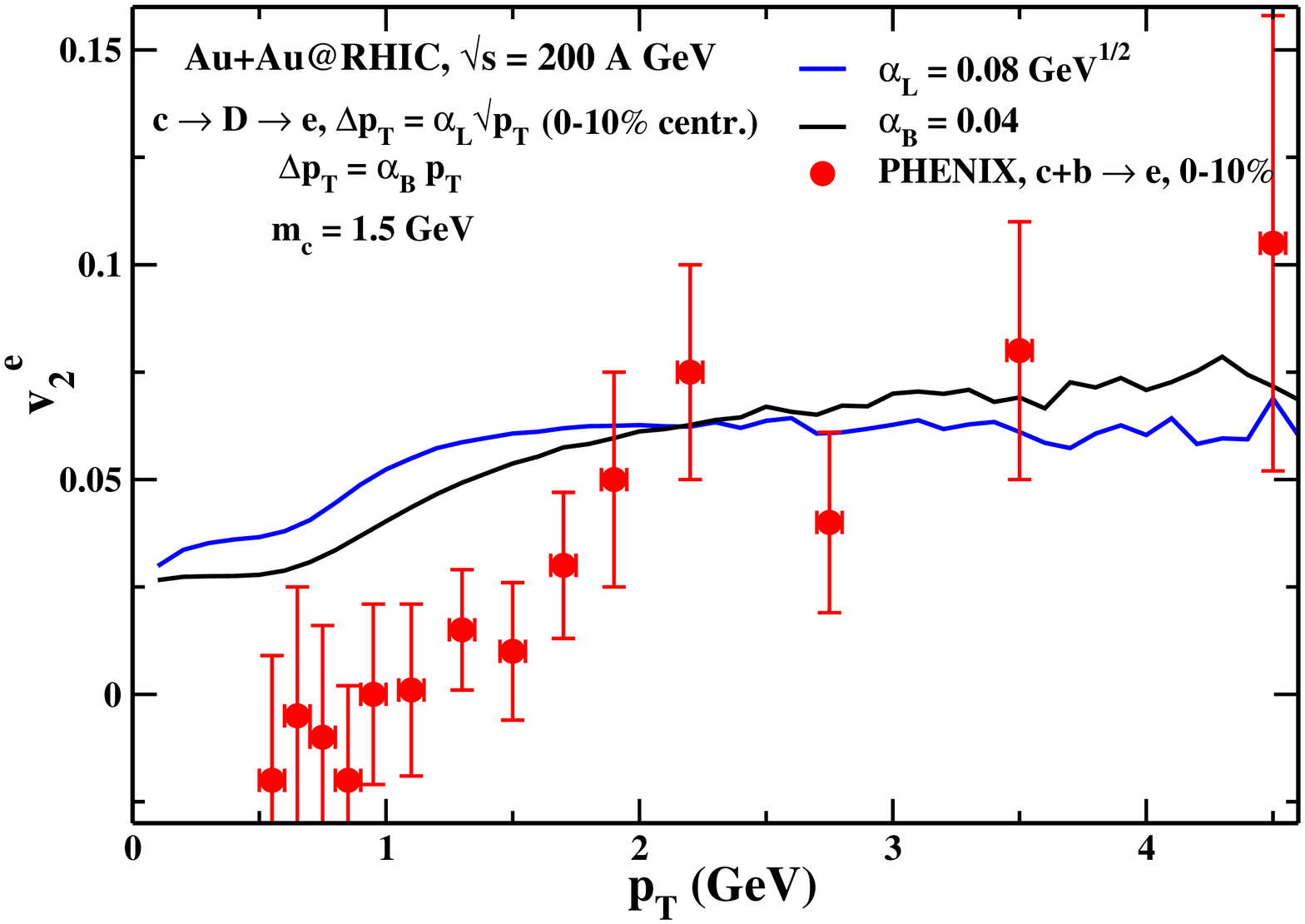}
\caption{(Colour on-line) $R_{AA}$ and $v_2$ for non photonic electrons at RHIC.
Top: Momentum loss/per collision $\propto$ momentum (left) and $\propto$
square-root of momentum (right). Bottom: Momentum loss per collision=
constant (left) and $v_2 (p_T)$ for single electrons.}
\label{fig1}
\end{center}
\end{figure*}

\section{Formulation}

\subsection{Charm Production}
The energy loss of quarks and gluons is most easily
seen via the suppressed production of hadrons measured
using nuclear modification factor, $R_{AA}$:

\begin{equation}
R_{AA}(p_{T},y)=\frac{dN_{AA}/d^{2}p_{T}dy}
{\langle T_{AA}\rangle d\sigma_{pp}/d^{2}p_{T}dy}
\label{Raa}
\end{equation}
where $N_{AA}$ is the hadron production for the nucleus-nucleus system at
a given impact parameter, $T_{AA}$ is the corresponding nuclear thickness,
and $\sigma_{pp}$ is the cross-section for the production of hadrons at the
corresponding centre of mass energy/nucleon in $pp$ collisions.

One can now calculate
\begin{eqnarray}
\frac{d\sigma_{pp}}{dy_1 dy_2 d^{2}p_{T}} &=& 2 x_{a}x_{b}\sum_{ij}
\left[f^{(a)}_{i}(x_{a},Q^{2})f_{j}^{(b)}(x_{b},Q^{2})
\frac{d\hat{\sigma}_{ij}(\hat{s},\hat{t},\hat{u})}{d\hat{t}}
\right.\nonumber\\
&+& \left.f_{j}^{(a)}(x_{a},Q^{2})f_{i}^{(b)}(x_{b},Q^{2})
\frac{d\hat{\sigma}_{ij}(\hat{s},\hat{u},\hat{t})}{d\hat{t}}\right]
/(1+\delta_{ij})~,
\label{sigma}
\end{eqnarray}
where $p_{T}$ and $y_{1,2}$ are the momenta and rapidities of
produced charm and anti-charm and
$x_{a} $ and $x_{b} $ are the fractions of the momenta carried by the partons
from their interacting parent hadrons.
These are given by
\begin{equation}
x_{a}=\frac{M_{T}}{\sqrt{s}}(e^{y_1}+e^{y_2})~;~~~~
x_{b}=\frac{M_{T}}{\sqrt{s}}(e^{-y_1}+e^{-y_2})~.
\label{x}
\end{equation}
where $M_{T} $ is the transverse mass, $\sqrt{m_{Q}^{2}+p_{T}^{2}}$,
of the produced heavy quark.
The subscripts $i$ and $j$ denote the interacting partons, and $f_{i/j}$
are the partonic distribution functions for the nucleons.
The fundamental processes included for LO calculations are:

\begin{eqnarray}
g+g \rightarrow c+\overline{c}\nonumber\\
q+\bar{q} \rightarrow c+\overline{c}~.
\label{process}
\end{eqnarray}

We recall that the above LO pQCD expression reproduces the NLO
results~\cite{MNR}
when supplemented with a $K$-factor $\approx$ 2 (see Ref.~\cite{jamil}).

We have used  $T_{AA}$= 225 fm$^{-2}$ for 0-10$\%$ centrality
for Au+Au
collisions at RHIC, as calculated from Glauber
formalism. For Pb+Pb collisions at LHC,
$T_{AA}$ =195 fm$^{-2}$ for 0-20$\%$ centrality has been used. We use CTEQ5M
structure function along with EKS98~\cite{EKS98} shadowing function.
 The factorization,
renormalization, and fragmentation scales are chosen as $\sqrt{m_Q^2+p_T^2}$
and the charm quark mass has been taken as 1.5 GeV.

\subsection{\bf {Energy Loss}}
We propose an empirical model for the energy loss for charm quarks
which is inspired by a multiple scattering model used earlier by~\cite{huang}
supplemented with considerations of Baier et al~\cite{bdmps} for partonic
energy loss~\cite{jeon}.
 A straight-forward empirical implementation of
a similar energy loss procedure for light parton
has been found to provide a satisfactory dependence of energy loss
on the centralitiy of the collision~\cite{somnath}.
 We recall
once again that the nuclear modification of heavy meson production is similar
to those for light mesons.

We perform a Monte Carlo implementation of our model
calculations and estimate the momentum loss
of charm quarks and nuclear modification of D-meson
and single electron production.
We assume that the energy loss of heavy quarks proceeds via multiple
collisions and that the momentum loss per collision is given by,
(see for example Ref.~\cite{muller})
\begin{equation}
 (\Delta p)_i=\alpha \, (p_i)^{\beta}~,
 \label{deltapt}
\end{equation}
so that one can write
\begin{equation}
\frac{dp}{dx}=-\frac{\Delta p}{\lambda}
\label{dpdx}
\end{equation}
where $\alpha$ and $\beta$ are parameters to be determined
and $\lambda$ is the mean free path of the charm quark, taken as 1 fm, in
these initial studies.
In what follows, we shall consider charm quarks at central
rapidities and thus $p=p_T$. Thus the momentum
of the charm quark after $n$ collisions will be given by
\begin{equation}
p_{n+1}=p_n-(\Delta p)_n
\end{equation}
The charm quark can continue to lose energy in collisions as long as
the resulting momentum remains positive.
We estimate the probability for the charm quark to have $n$ collisions,
 while covering the path length $L$  from a Poisson distribution
\begin{equation}
P(n,L)=\frac{(L/\lambda)^{n}}{n!}e^{-L/\lambda}.
\label{prob}
\end{equation}

 Taking a value for the coefficient $\alpha$
and the exponent $\beta$, we estimate the largest number
of collisions- $N$, which the charm quark
having momentum $p_T$ can undergo. Next we sample the number of
collisions $n$, which the charm undergoes from the distribution
\begin{equation}
p(n)=P(n,L)/\sum_{n=1}^N P(n,L)
\end{equation}
to get the final momentum of the charm quark.

Finally we fragment the charm quarks
into D-mesons. Thus we have,
\begin{equation}
E\frac{d^3\sigma}{d^3p}=E_Q\frac{d^3\sigma(Q)}{d^3p_Q}\otimes
D(Q\rightarrow H_Q)\otimes F(H_Q\rightarrow e)~,
\label{d-meson}
\end{equation}
where  the fragmentation of the heavy quark $Q$ into the heavy-meson $H_Q$ is
described by the fragmentation function $D$. We have assumed that the shape of
$D(z)$, where $z=p_D/p_c$, is identical for all the $D$-mesons,~\cite{Peterson}
and
\begin{equation}
D^{(c)}_D(z)=\frac{n_D}{z[1-1/z-\epsilon_p/(1-z)]^2}~,
\label{frag}
\end{equation}
where $\epsilon_p$ is the Peterson parameter and
\begin{equation}
\int_0^1 \, dz \, D(z)=1~.
\end{equation}
We have kept it fixed at $\epsilon_p$=0.13.

Here $F(H_Q\rightarrow e)$ denotes semileptonic decay of $D$-mesons
and the electron distribution is taken from Ref.~\cite{Altarelli}.

\subsection{\bf {Azimuthal Anisotropy}}
Non-central collisions of identical nuclei will lead to
an oval overlap zone, whose length in and out of the reaction plain
would be different. Thus, charm quarks traversing the QGP in and out
of the plain will cover different path lengths and lose differing
amount of energy. This would lead to an azimuthal dependence in the
distribution of resulting charm mesons, whose azimuthal anisotropy
could be measured in terms of the $v_2$ coefficient defined by
\begin{equation}
v_{2}(p_T)=\frac{\int{d\phi \frac{dN}{p_{T}dp_{T}d\phi}\cos{(2\phi)}}}{\int{d\phi \frac{dN}{p_{T}dp_{T}d\phi}}}
\label{v2}
\end{equation}
We have approximated the colliding nuclei as having a uniform
density with radius
$R$ in these calculations and obtained average path-length for the charm quarks
along a given $\phi$ using Eq.~9 of Ref.~\cite{somnath}.

\begin{figure*}[h]
\begin{center}
\includegraphics[height=3in,width=3in,angle=270]{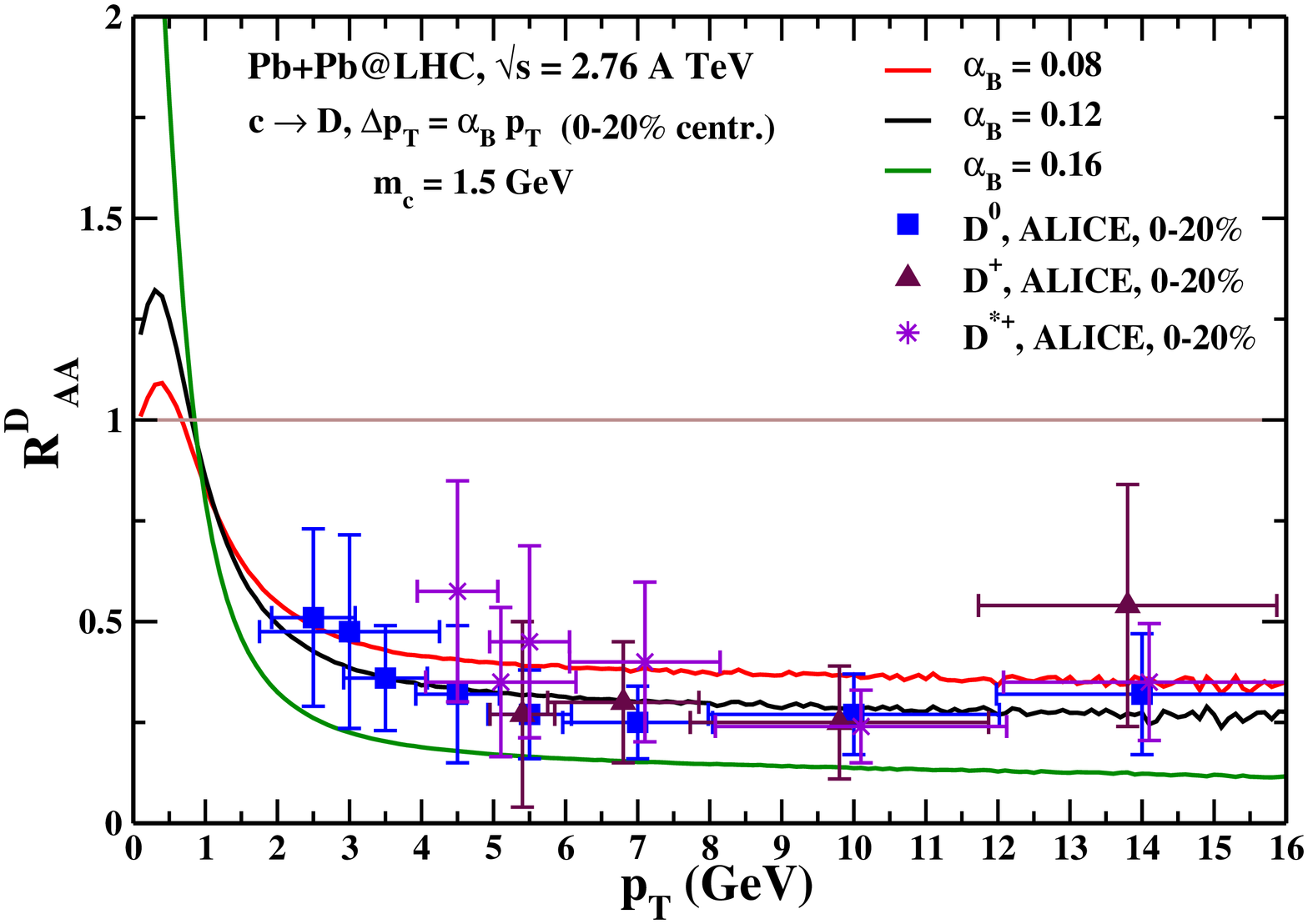}
\includegraphics[height=3in,width=3in,angle=270]{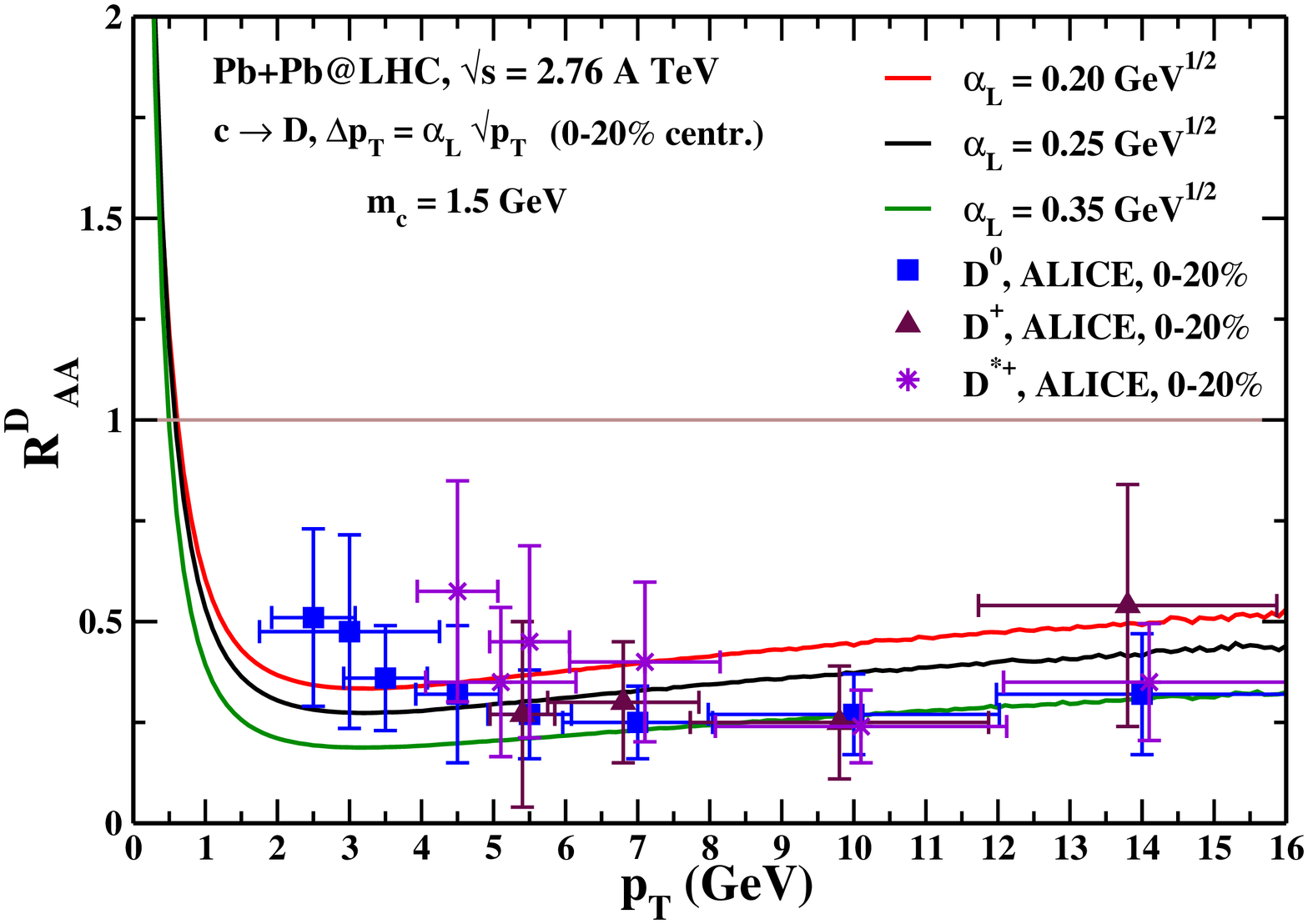}
\includegraphics[height=3in,width=3in,angle=270]{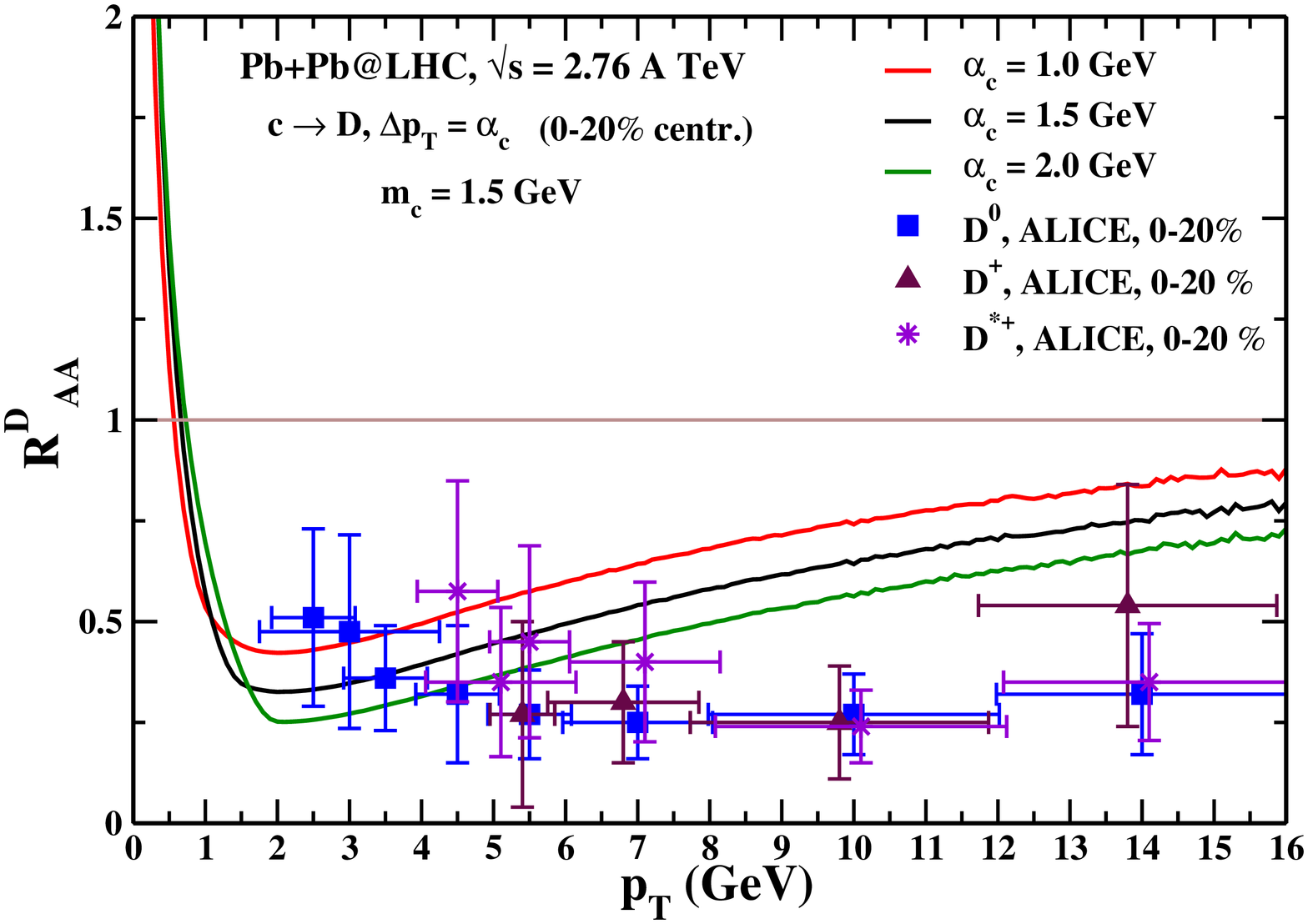}
\includegraphics[height=3in,width=3in,angle=270]{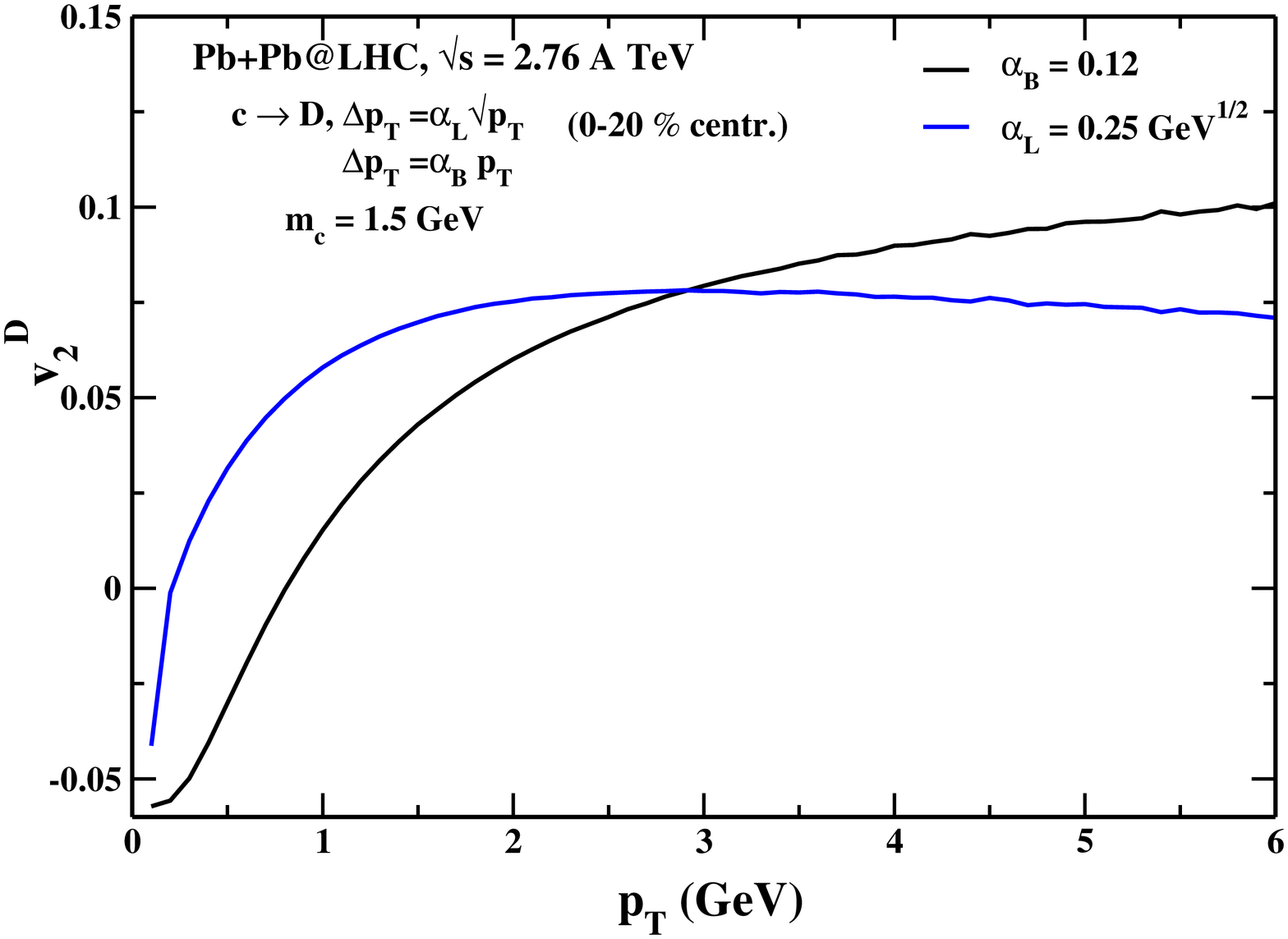}
\caption{(Colour on-line) $R_{AA}$ and $v_2$ for D mesons at LHC.
Top: Momentum loss per collision $\propto$ momentum (left) and
$\propto$ square root of momentum (right). Bottom: Momentum loss
per collision= constant (left) and $v_2 (p_T)$ for D-mesons (right).}
\label{fig2}
\end{center}
\end{figure*}

\begin{figure*}[h]
\begin{center}
\includegraphics[height=3in,width=3in,angle=270]{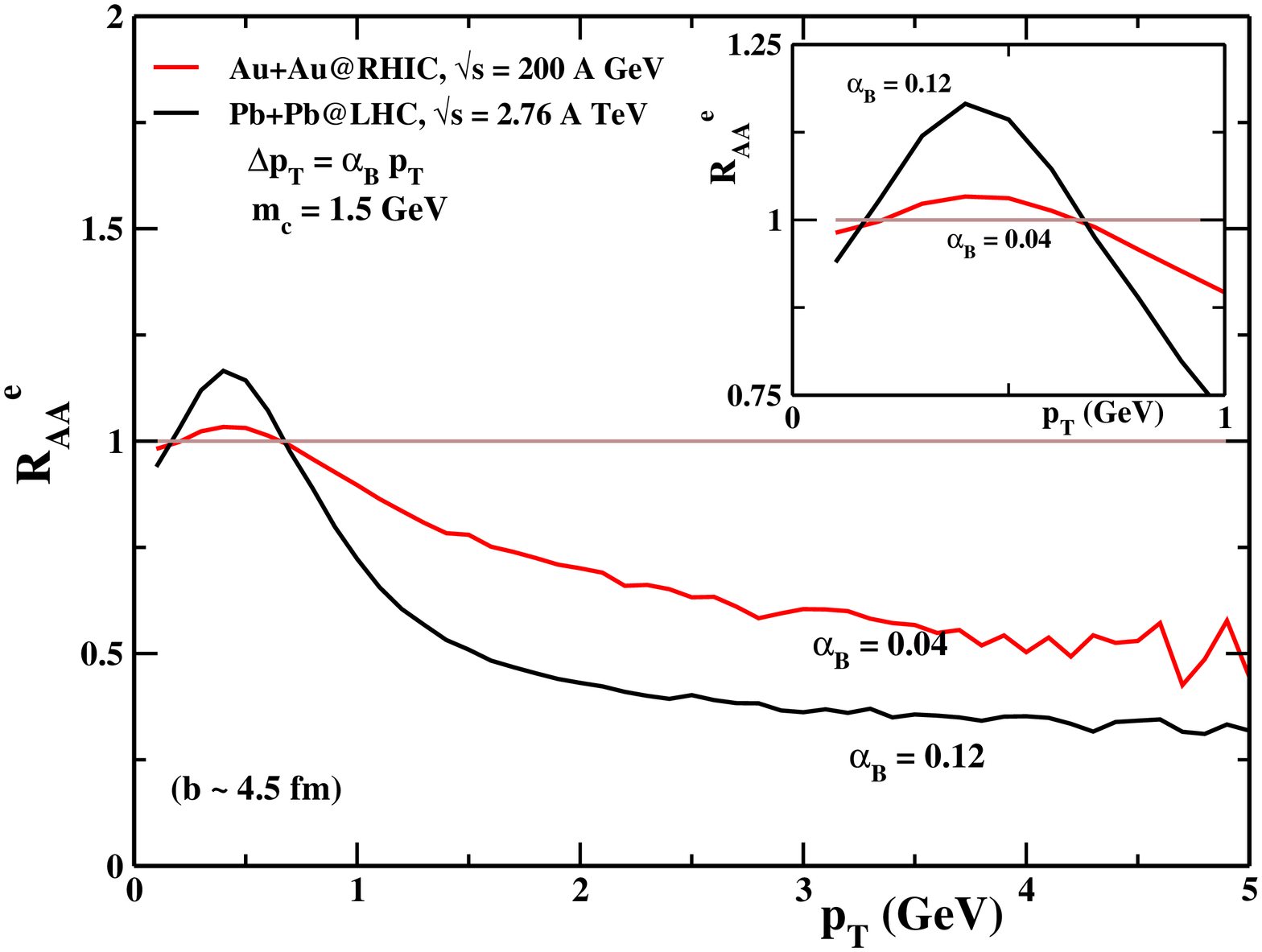}
\includegraphics[height=3in,width=3in,angle=270]{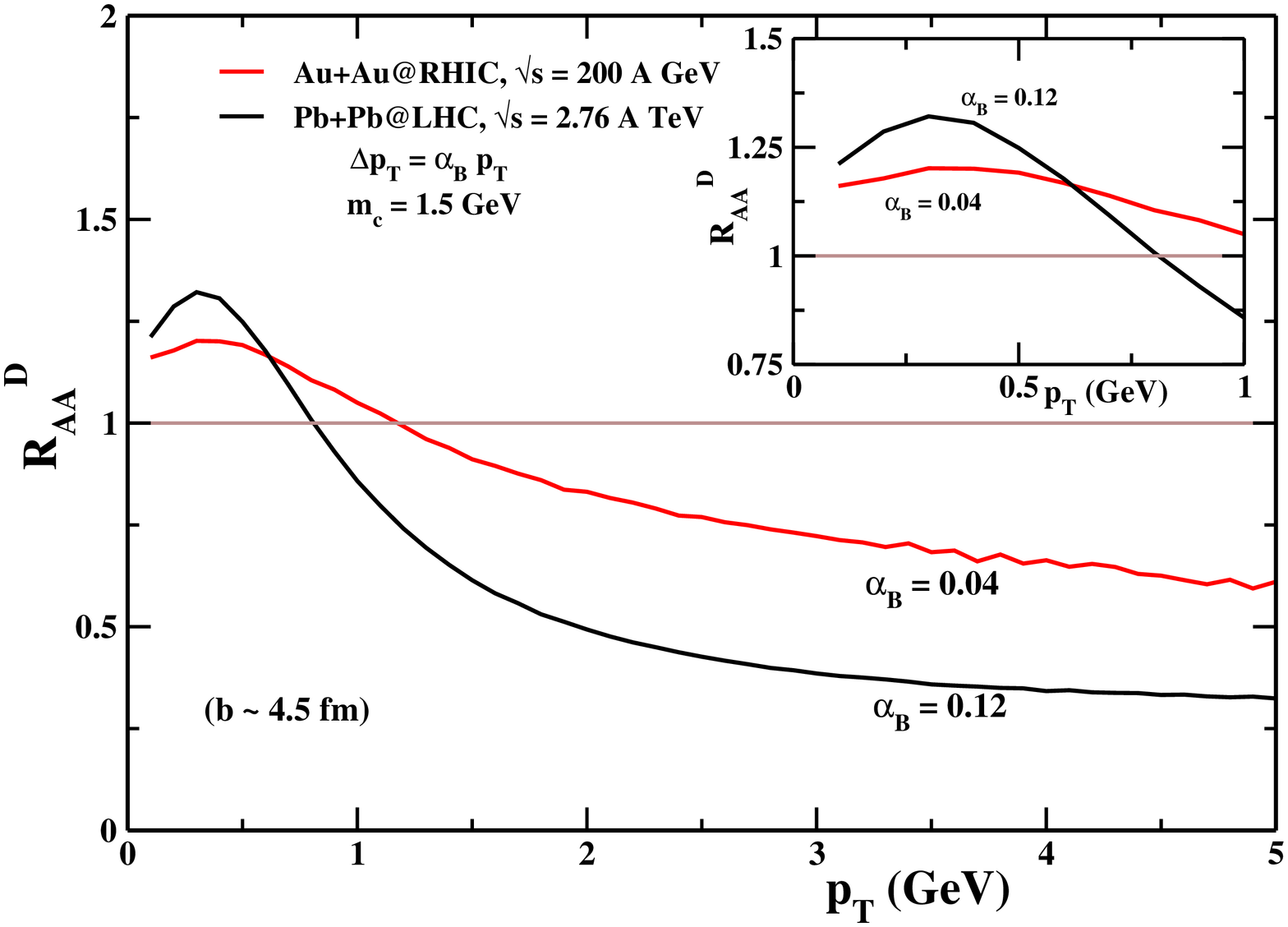}
\caption{(Colour on-line) Left: $R_{AA}$ for single electrons for
 collisions of
 gold nuclei at RHIC and lead nuclei at LHC at $b=$ 4.5 fm. Inset shows
enhanced production of single electrons having low $p_T$.
Right: $R_{AA}$ for D-mesons for the same collisions at
RHIC and LHC and the inset shows the enhanced production of D-mesons
having low $p_T$. The best fit values of $\alpha$ from Figs.~\ref{fig1}
and \ref{fig2} and $\beta$=1 are used for these calculations.}
\label{fig3}
\end{center}
\end{figure*}

\section{\bf{Results and Discussion}}

Let us now discuss our results for nuclear modification
factor and azimuthal anisotropy.
We consider three values for the exponent $\beta$;
0, 0.5, 1.0 appearing in Eq.~\ref{deltapt}, inspired by
the three energy loss mechanisms, namely those applicable
in the so-called Bethe-Heitler regime, LPM regime, and complete
coherence regimes considered by Baier et al.~\cite{bdmps,somnath}
which lead to energy loss per unit length as proportional to
energy, square-root of the energy, and independent of the
energy for light partons.  Kampfer et al~\cite{kaempfer} had earlier
used this approach to study the effect of charm quark energy loss on
the correlated charm decay.

Next we vary $\alpha$ to get a description of the $R_{AA}$ for
single electrons  at RHIC (Fig.~\ref{fig1}) and for $D$-mesons
at LHC (Fig.~\ref{fig2}).

In Fig.~\ref{fig1} we show our results for nuclear modification
factor $R_{AA}$ and azimuthal anisotropy $v_2$ for
non photonic electrons at top RHIC energy $\sqrt{s}$ =
200 A GeV~\cite{phenix1}.
 Comparing the results of Figs.~\ref{fig1}, we see that
the model assuming momentum loss per collision as proportional to
the momentum
closely follows the shape of the
 experimentally determined $R_{AA}$ for single electrons almost over the
entire range of $p_T$ under consideration. We add that our theoretical
calculations have not included the $b \rightarrow e$ contribution which is
contained in the experimental results which can modify the $R_{AA}$ for
larger $p_T$ by up to 10\% as the produced $b$ quarks are much less
in number and also lose much smaller energy~\cite{bottom}. The
scenario, where $\Delta p \propto \sqrt{p}$, is only moderately successful
in describing the data over a limited $p_T$ range of 2--4 GeV/$c$
(Fig.~\ref{fig1}). While the assumption of a constant momentum loss
per collision may bracket the $R_{AA}$ over the very limited range of 2--3
GeV/$c$, it does not follow the shape of the $p_T$ dependence
 (Fig.~\ref{fig1}).

The best values of the $\alpha$ determined from the results in
Fig.~\ref{fig1} are used to estimate $v_2$ for the single electrons.
 We see that our calculations provide a reasonable
description of $v_2(p_T)$ for $p_T \geq$ 2 GeV/$c$ and overestimate the
results for lower $p_T$. A relaxation of our
 assumption of a static medium at a constant temperature and
a uniform density of the nuclei may improve this agreement.

Similar results are obtained when we apply the model to the $R_{AA}$
measured~\cite{alice1} for the $D$-mesons at the LHC (Fig.~\ref{fig2}).
We again see that the
model using $\Delta p \propto p$ provides a good description of the data
over the entire $p_T$ range, while that using
$\Delta p \propto \sqrt{p}$ seems to describe the data for
 $p_T \geq$ 4 GeV/$c$. The constant momentum
transfer collision misses the shape of the
$p_T$ distribution completely though it is able to bracket the
numerical values over a very narrow $p_T$ range of 3--5 GeV/$c$
(Fig.~\ref{fig2}). The predictions for $v_2$ are given
for a ready reference.

Several factors could affect the value of the energy loss
coefficient '$\alpha$' for a given mechanism. We have verified
that increasing (decreasing) the mean free path, $\lambda$, by
0.5 fm results in a decrease (increase) of the coefficient, $\alpha_B$
such that $\alpha_B/\lambda$ remains unaltered.

We have kept $\alpha_s$ fixed at 0.3 while estimating the initial
charm distribution. Taking the renormalization scale as
$C\sqrt{p_{T}^{2}+M_{Q}^{2}}$, with $C$= 1 or 2 leads to a decrease
in the value of $R_{AA}$ by 7-10 \%, which can then be offset by
decreasing $\alpha_B$ by about 12 \%.

It is interesting to recall that a Focker-Planck equation
given by
\begin{equation}
\frac{\partial{f}}{\partial{t}}=\frac{\partial}{\partial{p_i}}\left[A_{i}(\textbf{p})f
+\frac{\partial}{\partial{p_j}}(B_{ij}(\textbf{p})f)\right]~,
\label{focker}
\end{equation}
describes the evolution of the distribution '$f$', of charm quarks
propagating in quark gluon plasma~\cite{svetitsky}
and losing energy due to multiple
soft scatterings with light quarks and gluons. This
leads to a drag, $A_{i}(\textbf{p})$, and a diffusion $B_{ij}(\textbf{p})$
on the momentum of the charm quark.
Assuming that $A_{i}(\textbf{p})$ depends on momentum only, we have
\begin{equation}
A_{i}(\textbf{p})=A(p^2)\,p_i~,
\label{drag}
\end{equation}
and the energy loss $dE/dx$ can be related to drag
coefficient $A(p^2)$ by
\begin{equation}
\frac{dE}{dx}=-A(p^2)\, p~,
\end{equation}
where $E$ is the energy of the charm quark, and
$p$ its momentum. Considering the average temperature of the plasma
attained at RHIC as $\approx$ 220 MeV~\cite{Phenix_photon}, we can
read the drag coefficient from Fig.~3(a) of Ref.~\cite{munshi} as $\approx$
0.02 fm$^{-1}$ for $p_T$ up to 5 GeV/$c$.
We can re-write the above equation as
\begin{equation}
\frac{dp}{dx}=-A(p^2)\,E~,
\label{eloss1}
\end{equation}
Comparing this with one of the ansatzes used for
energy loss per collision in the present work, namely
$\Delta p=\alpha_B~ p$, we can write that
\begin{equation}
\frac{dp}{dx}=-\frac{\alpha_B}{\lambda}\,p\ ,
\label{eloss2}
\end{equation}
Thus the effective drag coefficient '$A_{\rm{eff}}(p^2)$',
can be written as
\begin{equation}
A_{\rm{eff}}(p^2)=\frac{\alpha_B}{\lambda}~\frac{p}{E}
\end{equation}
which reduces to $\alpha_B/\lambda$ for large values of $p$

We thus note that the effective drag at RHIC energies is about
0.04 fm$^{-1}$ compared to 0.02 fm$^{-1}$ estimated for soft multiple collisions
by authors of Ref.~\cite{munshi}. Thus we conclude that at RHIC energies only
half of the energy loss could be due to collisions, while the other half could
attributed to radiations of gluons. Similarly estimating the average temperature
at LHC as about 270 MeV and using the results for drag due to collisions as
$\approx$0.04 fm$^{-1}$ from Ref.~\cite{munshi} at high momentum, we note that
the collisions account for only one-third of energy loss at 2.76 TeV/nucleon.

It is of interest to compare our results with other studies on medium
modification of charm propagation reported in the literature. Thus
Moore and Teaney~\cite{Moore} have calculated the diffusion '$D$' and
drag coefficient '$\eta_D$', (denoted by '$A$' here) using
LO pQCD as
\begin{equation}
D\approx\frac{6}{2\pi T}\left(\frac{0.5}{\alpha_s}\right)^2
\label{DMoore}
\end{equation}
and
\begin{equation}
\eta_D=\frac{T}{M_{Q}D}
\label{nMoore}
\end{equation}
Taking $\alpha_s\,\approx$0.3, this provides $\eta_D\,\approx$0.06 fm$^{-1}$
at 220 MeV and $\eta_D\,\approx$0.09 fm$^{-1}$ at 270 MeV, which are larger than
our values at RHIC and smaller than those at 2.76 TeV/nucleon at LHC.

Recall however, the results of Bass et al.~\cite{Bass2} that a description of
medium modification as well as $v_2$ for single electrons at RHIC brackets
the diffusion coefficient '$D$' between $1.5/2\pi T$ and $6.0/2\pi T$, when flow
and contributions of bottom electrons is included. This provides a large value for the drag
coefficient between 0.17 and 0.68 fm$^{-1}$. At first these large values may look surprising.
However from Eq. \ref{DMoore} we see that these would correspond to values of $\alpha_s$
between $\approx$1.0 and $\approx$0.5, which are rather large and make the use of perturbative
QCD questionable.

In this connection the studies of Gossiaux et al~\cite{gossiaux}, are also of considerable interest which suggest
that the collisional energy loss could be substantially larger if the Debye mass is replaced
by hard thermal loop calculation and a running coupling constant is used. We further recall
the work of authors of Ref.~\cite{hees}, which suggests that a considerable drag could be produced
by the resonant heavy quark-light quark interaction, beyond that determined by LO pQCD
interactions.

We have already mentioned that $c$ quarks will materialize
mostly as $D$-mesons. This
can have an interesting consequence which
is already apparent in the Figs.~\ref{fig1}
and \ref{fig2}. The $c$ quarks after losing energy will pile
up at lower energies and
this would result in a characteristic increase in the production
 of $D$-mesons as
well as single electrons having lower transverse momenta (Fig.~\ref{fig3}).
We note that while
there is suppression by almost a factor of 2--4 depending on the
incident energy (for single electrons) at larger $p_T$, there is essentially
no suppression at lower $p_T$ at RHIC and even an increase by a few percent
at LHC where the energy loss is higher and the momentum spectra of $c$ quarks
have less steeper slopes.
The increase in the case of $D$-mesons at LHC is rather
spectacular.   This is different from the
normal enhancement of mesons having low $p_T$ due to the
so-called Cronin effect,
which is expected to be less important at higher incident energies
and heavy mesons (see e.g., Ref.~\cite{cronin}). We recall
that a similar enhancement in D-meson production
at low $p_T$ has been predicted by considering the drag suffered
the heavy quarks in the plasma and during their hadronization~\cite{rainer}.
In a forthcoming
paper we shall show that this can lead to a slight increase
 in the production of
low mass dileptons due to correlated charm decay~\cite{mdy4}.

At the very minimum the present work describes a simple procedure to implement
energy loss of heavy quarks in relativistic collision of heavy nuclei. It
will be of interest to explore the energy and centrality dependence of
the momentum loss coefficient $\alpha$.

\section{Summary}
We have used a simple model where charm quarks traversing QGP lose energy in
multiple collisions. Exploring different parametric forms of energy loss
we find that the form where the momentum loss per collision is proportional
to the momentum gives a good description of nuclear modification of single
electron production at RHIC and D-meson production at LHC. Comparing our results
with drag coefficients due to collisional energy loss calculated earlier, we find that
only a part of the energy loss could be due to collisions. We note a characteristic
increase in production of single electrons and D-mesons having low momenta
due to energy loss of charms having larger energy.

\section*{Acknowledgments}
One of us (MY) acknowledges financial support of the
Department of Atomic Energy, Government of India
during the course of these studies.

\end{document}